\documentclass[12pt,preprint]{aastex}

\newcommand{\ez}{\mbox{$\hat{{\bf e}}_{\rm z}$}}

\shorttitle{Damping of fast MHD oscillations in filament threads}
\shortauthors{Arregui et al.}

\begin{document}

\title{Damping of Fast Magnetohydrodynamic Oscillations\\ in 
Quiescent Filament Threads}

\author{I\~nigo Arregui\altaffilmark{1},  
Jaume Terradas\altaffilmark{1,2}, Ram\'on Oliver\altaffilmark{1}, and Jos\'e Luis
Ballester\altaffilmark{1}}

\altaffiltext{1}{Departament de F\'{\i}sica, Universitat de les Illes Balears,
E-07122 Palma de Mallorca, Spain. Email: inigo.arregui@uib.es, jaume.terradas@uib.es, 
ramon.oliver@uib.es, joseluis.ballester@uib.es}
\altaffiltext{2}{Centre Plasma Astrophysics, Katholieke Universiteit Leuven,
Leuven, B-3001, Belgium.}

\begin{abstract}
High-resolution observations provide evidence about the existence of small-amplitude 
transverse oscillations in solar filament fine structures. These oscillations
are believed to represent fast magnetohydrodynamic (MHD) waves and 
the disturbances are seen to be damped in short timescales of the order of 1 to 4 periods. 
In this Letter  we propose that, due to the highly inhomogeneous nature of the 
filament plasma at the fine structure spatial scale, the phenomenon of resonant absorption is likely 
to operate in the temporal attenuation of fast MHD oscillations. 
By considering transverse inhomogeneity in a straight flux tube model we find that, 
for density inhomogeneities typical of filament threads, the decay times are of
a few oscillatory periods only.
\end{abstract}


\section{Introduction}\label{intro}


Quiescent solar filaments form along the inversion polarity line or between the weak remnants 
of active regions. Early filament observations (\citealt{Engvold98}) as well as 
recent high-resolution H$\alpha$ observations, obtained with the Swedish Solar Telescope (SST) in 
La Palma (\citealt{Lin05}) and the Dutch Open Telescope (DOT), have revealed that their fine
structure is composed by many horizontal and thin dark threads.
The measured average width of resolved 
threads is about $0.3$ arc sec ($\sim$ $210$ km) while the length is between $5$ and $40$ 
arc sec ($\sim$ 3500 - 28,000  km). They seem to be partially filled with cold plasma
\citep{Lin05}, typically two orders of magnitude denser than that of the corona, and 
it is generally assumed that they outline their magnetic flux tubes
(\citealt{Engvold98,Linthesis,Lin05,Martin08}). This idea is strongly supported by the fact that they 
are inclined with respect to the filament long axis at a similar angle to what has been found for the 
magnetic field (\citealt{Leroy80,Bommier94,BL98}).


Small amplitude oscillations have been observed in filaments and it 
is well established that these periodic changes are of local nature.  
The detected peak velocity ranges from the noise level (down to 0.1 km s$^{-1}$ in 
some cases) to 2--3~km s$^{-1}$, although larger values have also been reported.  
Two-dimensional observations of filaments by \citet{YiEngvold91} and \citet{Yi91} revealed that
individual threads or groups of threads may oscillate independently with their own periods, 
which range between 3 and 20 minutes. More recently, \citet{Linthesis} reports that 
spatially coherent oscillations are found over slices, with an area of $1.4 \times 54$ arc sec$^2$, 
of a polar crown filament, and that among other, a significant periodicity 
at 26 minutes, strongly damped after 4 periods, appears.  Furthermore, \citet{Lin07} 
have shown evidence about traveling waves along a number of filament threads with an 
average phase velocity of $12$ \ km s$^{-1}$, a wavelength of $4''$ ($\sim 2800$ \ km), and
oscillatory periods of the individual threads that vary from $3$ to $9$ minutes. The observed periodic 
signals are obtained from Doppler velocity measurements and can therefore be associated to the transverse 
displacement of the fine structures.


The observed small amplitude oscillations  have been interpreted in terms of 
magnetohydrodynamic (MHD) waves (\citealt{OB02}) and theoretical models have been 
developed (see \citealt{Ballester05,Ballester06}, for recent reviews). 
\citet{diaz02} modeled a prominence thread as a straight cylindrical flux tube with a cool region 
representing the filament material, confined by two symmetric hot regions. They found that the fundamental fast mode 
is  always confined in the dense part of the flux tube, hence, for an oscillating cylindrical filament thread, 
it should be difficult to induce oscillations in adjacent threads, unless they are very close. 


The time damping of prominence oscillations has been unambiguously determined in some observations. 
Reliable values for the damping time have been derived, from different Doppler 
velocity time series, by \citet{Molowny-Horas99} and \citet{Terradas02}, in prominences, and 
by \citet{Linthesis}, in filaments.  The values thus obtained are usually between 
1 and 4 times the corresponding period, and large regions of prominences/filaments display similar damping times. The damping of perturbations is probably a common feature of filament oscillations, hence theoretical mechanisms must be explored and their damping time scales should be compared with those obtained from observations.  Linear non-adiabatic MHD waves have been proposed  as a potential mechanism to explain the observed attenuation time scales (\citealt{Carbonell04,Terradas01,Terradas05,Soler07,Soler08}).  Using thermal mechanisms only slow waves can be damped in an efficient manner while fast waves remain almost undamped.  Ion-neutral collisions provide a possible mechanism to damp fast waves (as well as Alfv\'en waves) that
is able to reproduce observed damping times for given parameter values, in particular for a quasi-neutral gas (\citealt{Forteza07}).


Apart from the mentioned non-ideal damping mechanisms, there is another possibility to attenuate 
fast waves in thin filament threads. The phenomenon of resonant wave damping is well documented  
for fast kink waves in coronal loops (see e.g.\ \citealt{GAA06,goossens08}, for recent reviews) and 
provides a plausible explanation to quickly damped transverse loop oscillations observed 
by TRACE (\citealt{Nakariakov99,Aschwanden99}). 
In this Letter, we address the resonant damping mechanism in the context of filament thread oscillations and assess 
its relevance in explaining the observed attenuation time scales.

\section{One-Dimensional Non-Uniform Filament Thread Model}

\begin{figure}
\epsscale{0.6}
\plotone{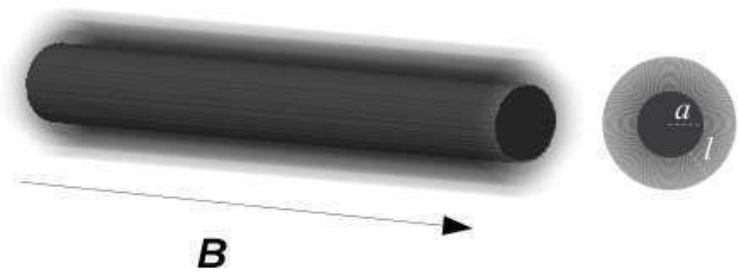} 
\caption{Sketch of the model used to represent a radially non-uniform filament fine structure of mean radius $a$, with a transverse inhomogeneity length-scale $l$ (see detail in the cross section).}
\label{fluxtube}
\end{figure}

Given the relatively simple structure of filament threads, when compared to the full
prominence/filament structure, the magnetic and plasma configuration of an individual and 
isolated thread can be theoretically approximated using a rather simplified model.
We consider a gravity-free, straight, cylindrically symmetric flux tube of mean radius 
$a$ (see Fig.~\ref{fluxtube}). In a system of cylindrical 
coordinates ($r$, $\varphi$, $z$) with the $z$-axis coinciding with the axis of the tube, the magnetic 
field is pointing in the $z$-direction, ${\bf B}=B\ez$. We neglect gas 
pressure, which allows us to concentrate on the oscillatory properties of fast and Alfv\'en MHD waves and 
their mutual interaction. In our straight field configuration this  zero-$\beta$ approximation
implies that the field strength is uniform and that the density profile can be 
chosen arbitrarily. The inhomogeneous filament thread is then modeled as a density enhancement with 
a one-dimensional non-uniform distribution of density, $\rho(r)$, across the structure. 
The internal filament plasma, with uniform density, $\rho_f$, occupies the full length of the 
tube and is connected to the coronal medium, with uniform density, $\rho_c$, by means of a 
non-uniform transitional  layer of thickness $l$. The ratio $l/a$ provides us 
with a measure of the transverse inhomogeneity length-scale, that can vary in between
$l/a=0$ (homogeneous tube) and $l/a=2$ (fully non-uniform tube).

\section{Damping of Linear Fast Kink Waves}

The magnetic flux tube model adopted here is a waveguide for a number of MHD oscillatory 
solutions. In the zero-$\beta$ approximation slow waves are absent. The properties of the 
remaining small amplitude fast and Alfv\'en waves can readily be described by considering 
the linear MHD wave equations for adiabatic changes of state for perturbations of the 
form $f(r)$exp$(\imath(\omega t+m\varphi-k_z z))$. Here $m$ and $k_z$ are the azimuthal and 
longitudinal wave-numbers and $\omega$ the oscillatory frequency. We further concentrate on 
perturbations with $m=1$, which represent fast kink waves that produce the transverse 
displacement of the tube as they propagate along the density enhancement.

\subsection{Analytical theory}
In the long wavelength or thin tube approximation ($k_za\ll1$) 
the frequency of the $m=1$ fast kink wave  can be written down analytically (see \citealt{ER83}) as 

\begin{equation}
\omega=k_z\sqrt{\frac{\rho_f V^2_{Af}+\rho_c V^2_{Ac}}{\rho_f+\rho_c}},
\end{equation}

\noindent
with $V_{Af,c}=B/\sqrt{\mu\rho_{f,c}}$  the filament (f) and coronal (c) Alfv\'en velocities. 
By defining $c=\rho_f/\rho_c$, for the density contrast, the period of kink oscillations with a 
wavelength $\lambda=2\pi/k_z$ can be written as 

\begin{equation}
P=\frac{\sqrt{2}}{2}\frac{\lambda}{V_{Af}}
\left(\frac{1+c}{c}\right)^{1/2}.\label{period}
\end{equation}

\noindent
The factor containing the density contrast varies between $\sqrt{2}$ and $1$, when $c$ is allowed 
to vary between a value slightly larger that $1$ (extremely tenuous thread) and $c\rightarrow\infty$.
For typical filament thread densities, which are two orders of magnitude larger than coronal densities, 
this factor is near unity. Equation~(\ref{period}) then predicts Alfv\'en velocities in the thread
as low as a few km s$^{-1}$ or as large as $\sim 200$ km s$^{-1}$, for  
combinations of periods and wavelengths  between 3 and 20 minutes, and $3000$  and $20,000$ km, respectively.

For fast kink waves to be damped by resonant absorption, transverse inhomogeneity in the Alfv\'en velocity
has to be considered. In our uniform field model this is obtained by considering $l\neq0$.  
Then, the $m=1$ solution is resonantly coupled to local Alfv\'en waves.
The coupling produces the temporal attenuation of fast transverse motions which are converted into localized
azimuthal Alfv\'enic oscillations. Asymptotic analytical expressions for the damping time, $\tau_d$, can be obtained 
under the assumption that the transverse inhomogeneity length-scale is small ($l/a \ll 1$). 
This is the so-called thin boundary approximation. When the long wavelength and thin boundary approximations are combined,
the analytical expression for the damping time over period can be written as  
(see e.g.\ \citealt{HY88,sakurai91,goossens92,goossens95,RR02})

\begin{equation}
\frac{\displaystyle \tau_{d}}{\displaystyle P} = F \;\; \frac {\displaystyle a}
{\displaystyle l}\;\; \frac{\displaystyle c + 1}{\displaystyle
c - 1}. \label{dampingrate}
\end{equation}

\noindent
Here $F$ is a numerical factor that depends on the particular variation of the density in the non-uniform layer. 
For a linear variation $F=4/\pi^2$ \citep{HY88,goossens92}; for a sinusoidal variation $F=2/\pi$ \citep{RR02}.  
For example, considering $c=200$ a typical density contrast and $l/a=0.1$ equation~(\ref{dampingrate}) predicts a damping time  
of $\sim$ 6 times the oscillatory period.

Figure~\ref{fig2} shows analytical estimates 
computed using equation~(\ref{dampingrate}) (solid lines). The damping is affected by the 
density contrast in the low contrast regime and $\tau_d/P$ rapidly decreases for increasing thread density 
(Fig.~\ref{fig2}a). Interestingly, it stops being dependent on this parameter in the large contrast regime, 
typical of filament threads. 
The damping time over period is independent of the wavelength of perturbations (Fig.~\ref{fig2}b), 
but rapidly decreases with increasing  inhomogeneity length-scale (Fig.~\ref{fig2}c).  
These results suggest that resonant absorption is a very efficient mechanism for the 
attenuation of fast waves in filament threads, especially because large thread densities and 
transverse plasma inhomogeneities can be combined together.

\subsection{Numerical results}

The two approximations used to derive equations~(\ref{period}) and (\ref{dampingrate}) 
may impose limitations to the applicability of the obtained results to filament thread oscillations. 
To start with, it is not clear how accurate the long wavelength approximation  can be, 
especially for short wavelengths. The period in equation~(\ref{dampingrate})
corresponds to the long wavelength limit (\ref{period}), which is independent of the radius of the structure, and does not include 
effects due to radial density inhomogeneity. Also, equation~(\ref{dampingrate}) 
is only valid as long as the damping time is sufficiently larger than the period, an assumption made in order to derive 
the expression, and clearly contradicted by the obtained results. The accuracy of the 
damping formula (\ref{dampingrate}) was assessed by \citet{tom04b} in the relatively low density contrast regime corresponding to damped coronal loop oscillations. Some degree of inaccuracy in the large contrast regime
characteristic of filament threads is also to be expected. 
We have therefore computed numerical approximations to the solutions by solving the full set of linear, resistive, 
small amplitude MHD wave equations for the $m=1$ transverse kink oscillations 
(see e.g.\ eqs.~[1]--[5] in \citealt{TOB06}), using the PDE2D code (\citealt{sewell05}). 

\begin{figure*}
\includegraphics[width=8cm,height=6cm]{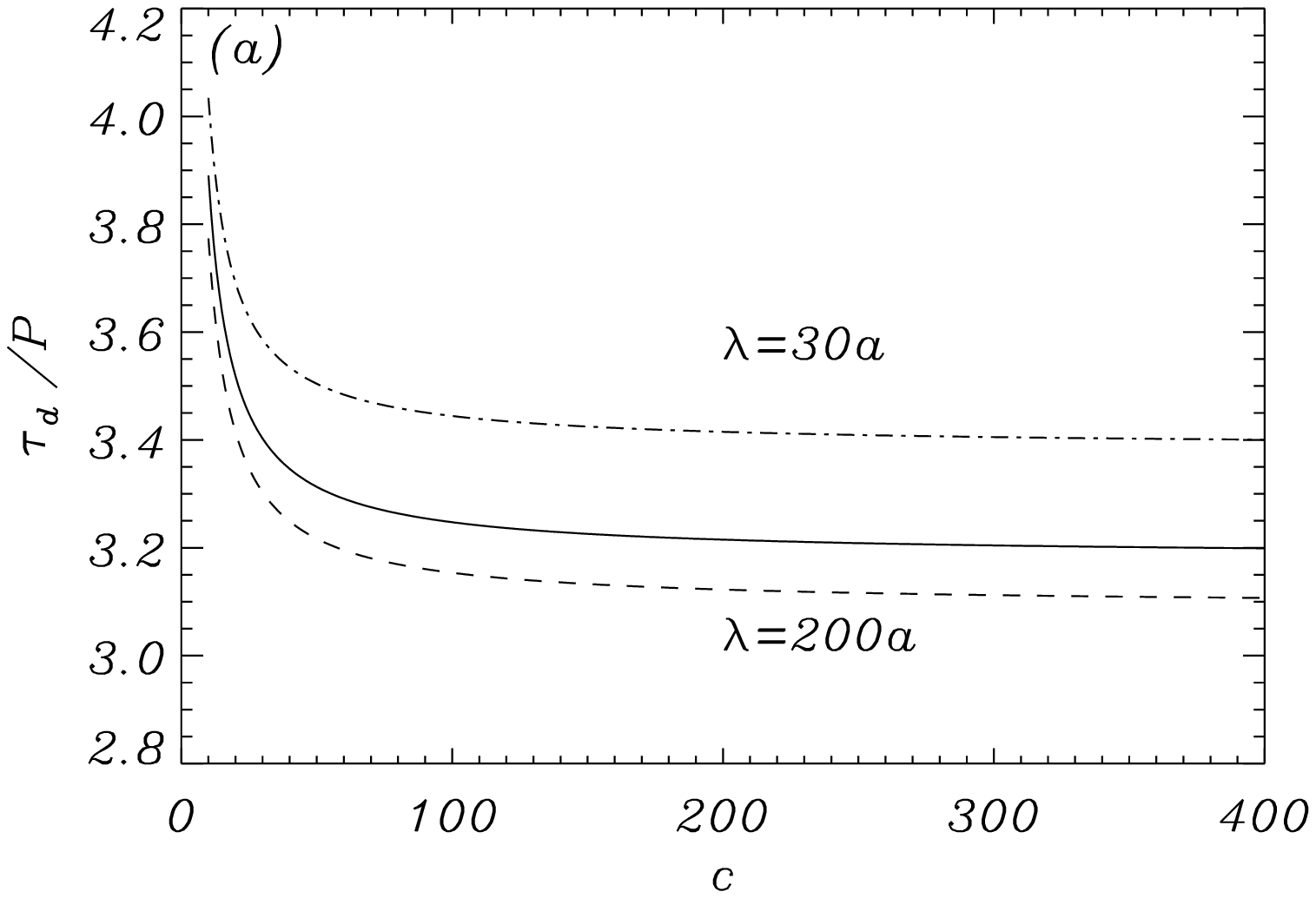}
\includegraphics[width=8cm,height=6cm]{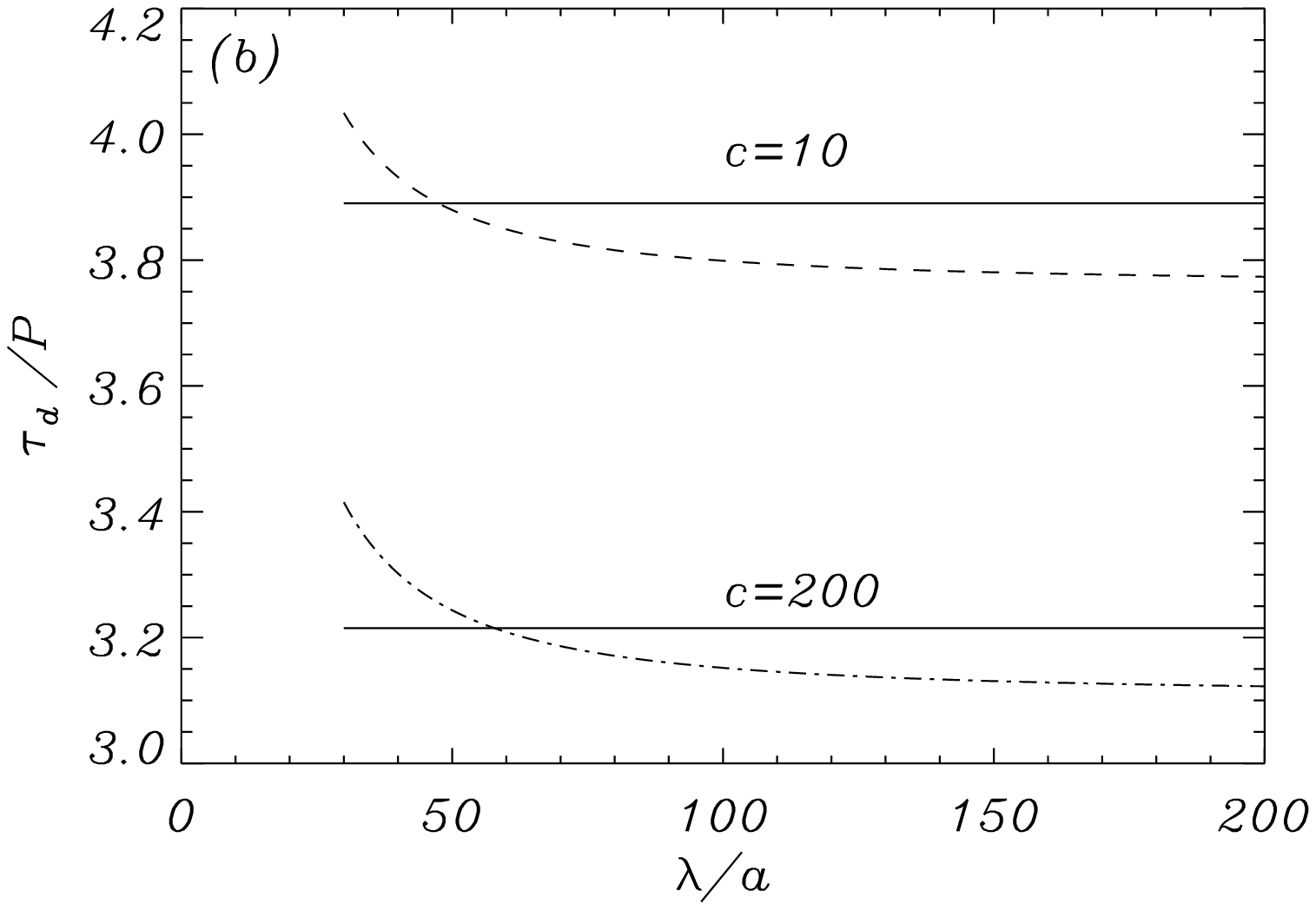}\\
\includegraphics[width=8cm,height=6cm]{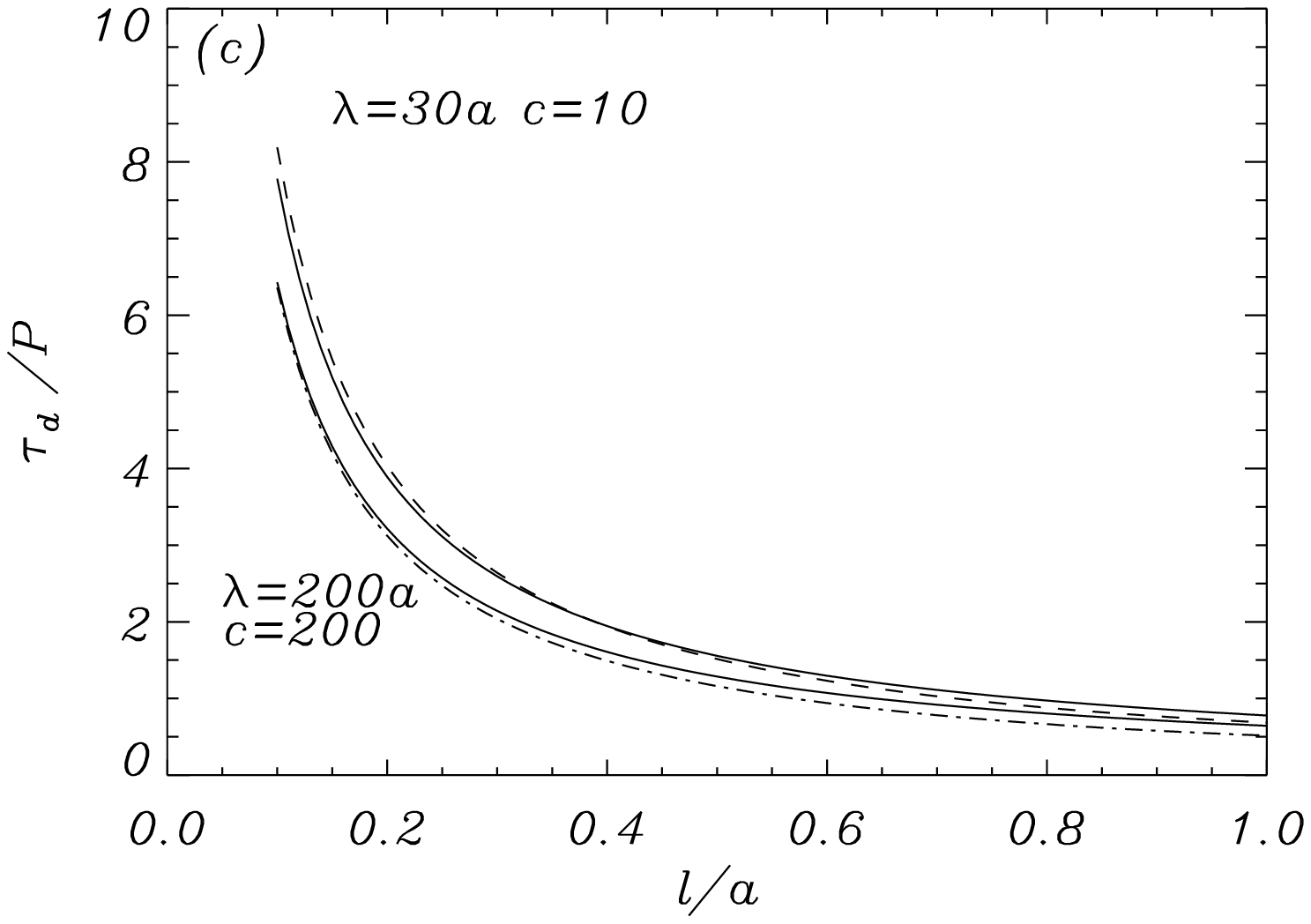}
\includegraphics[width=8cm,height=6cm]{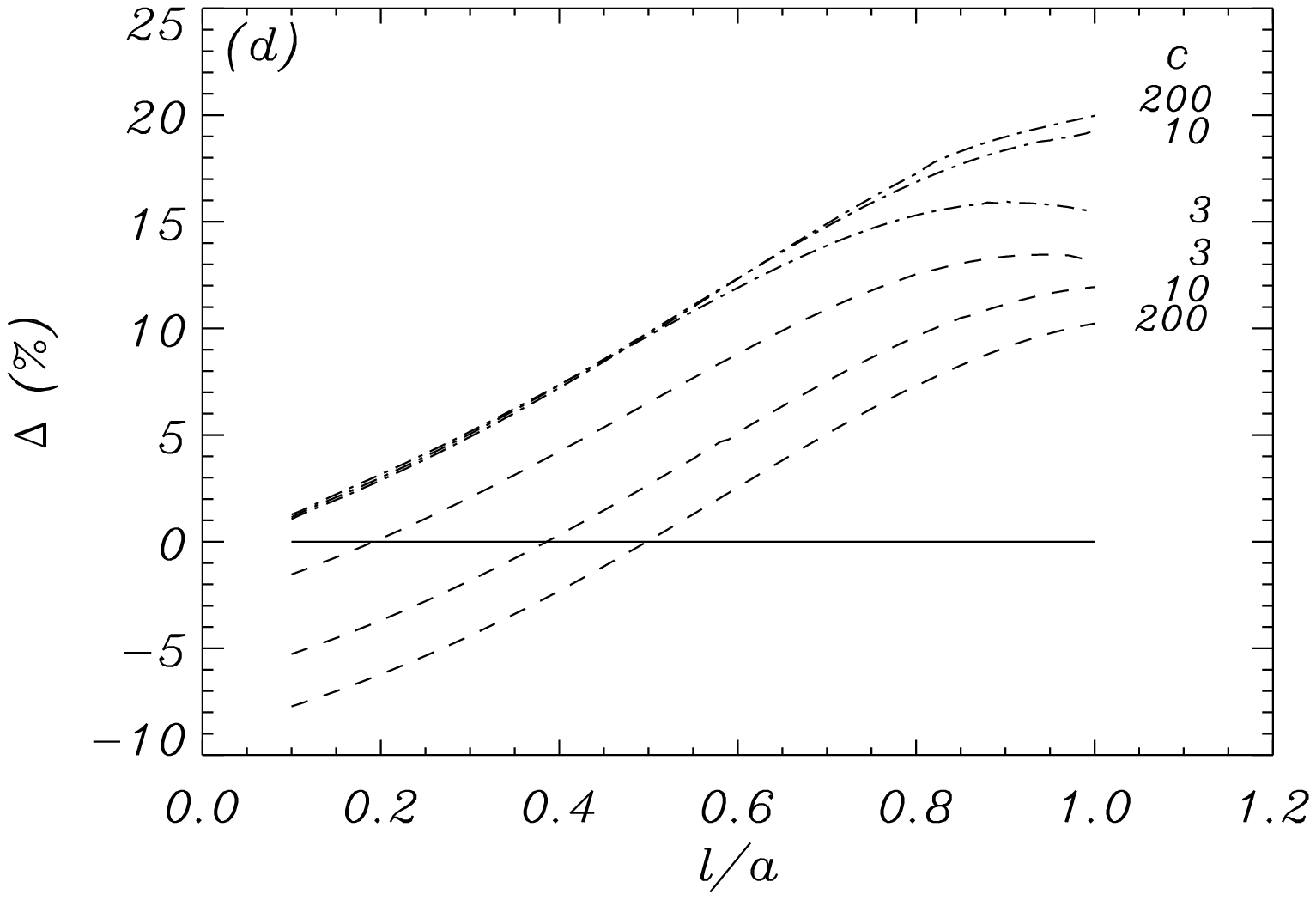}
\caption{{\em (a)--(c):} Damping time over period for fast kink waves in filament threads with $a=100$ km.
In all plots solid lines correspond to analytical solutions given by equation~(\ref{dampingrate}), with $F=2/\pi$.
{\em (a):} As a function of density contrast, with $l/a=0.2$ and for two wavelengths. {\em (b)}: 
As a function of wavelength, with $l/a=0.2$, and for two density contrasts. {\em (c)}: As a function of transverse inhomogeneity length-scale, for two combinations of wavelength and density contrast. {\em (d)} Percentage difference, $\Delta$, with respect to analytical formula (\ref{dampingrate}) for different combinations of wavelength, $\lambda=30a$ (dashed lines); $\lambda=200a$ (dash-dotted lines), and density contrast.}
\label{fig2}
\end{figure*}

Figure~\ref{fig2} shows the obtained numerical results. 
Analytical and numerical solutions display the same qualitative behavior with density contrast and transverse inhomogeneity length-scale (Figs.~\ref{fig2}a, c). Now the damping time over period slightly depends on the wavelength of perturbations (Fig.~\ref{fig2}b). Equation (\ref{dampingrate}) underestimates/overestimates this magnitude for short/long wavelengths.
The differences are small, of the order of 3\% for $c=10$, and do not vary much with density contrast for 
long wavelengths ($\lambda=200a$), but increase until 6\% for short ones ($\lambda=30a$). 
The long wavelength approximation is responsible for the discrepancies obtained for thin non-uniform layers 
(Fig.~\ref{fig2}c). Figure~\ref{fig2}d shows how  accurate equation~(\ref{dampingrate}) is  
for different combinations of wavelength, density contrast, and inhomogeneity length-scale.
For thin layers ($l/a=0.1$) the inaccuracy of the long wavelength approximation produces differences up to $\sim$ 10\% for the combination of short wavelength with high contrast thread. For thick layers, differences of the order of 20\% are obtained (in agreement with \citealt{tom04b}). Here, the combination of large wavelength with high contrast thread produces the largest discrepancy. Numerical results allow the computation of more accurate values, 
but do no change our previous conclusions regarding the efficiency and properties of resonant damping of transverse oscillations in filament threads.

\section{Discussion} \label{conclusion}

In this Letter we have shown that due to the highly inhomogeneous nature of filaments  at 
their transverse scales the process of resonant absorption is an efficient damping mechanism for 
fast MHD oscillations propagating in these structures. The relevance of the mechanism has been assessed 
in a flux tube model, with the inclusion of transverse inhomogeneity in the Alfv\'en velocity. Also,
the accuracy of analytical estimates, in terms of wavelength, density contrast, and transverse inhomogeneity, has been quantified.
For the typical large filament to coronal density contrast the mechanism produces rapid damping
in timescales of the order of a few oscillatory periods only. The obtained damping rates are only slightly 
dependent on the wavelength of perturbations. An important result is that the damping rate
becomes independent of density contrast for large values of this parameter. This has two seismological consequences.
First, the observational determination of density contrast is less critical than in the low contrast regime.
Second, according to seismic inversion results that combine theoretical and observed periods and damping times
(\citealt{Arregui07,GABW08}), high density thread models would be compatible with relatively short 
transverse inhomogeneity length-scales. Analytical estimates of $l/a\sim 0.15$  can even be calculated using 
equation~(\ref{dampingrate}) for a given observed $\tau_d/P=4$, taking the limit $c\rightarrow\infty$.

We believe our conclusion on the relevance of resonant damping in filament thread oscillations is robust in front 
of the main simplification adopted for the present study. Observations show that threads are only 
partially filled with cool and dense plasma and our one-dimensional model misses this property. 
Although the resonant damping mechanism relies on the Alfv\'en velocity inhomogeneity in the 
direction transverse to the magnetic field, an investigation of the damping properties in such a 
two-dimensional configuration is a necessary issue. 
Finally, the presence of flows is commonly observed along filament threads and the interplay 
between resonant damping and flows must also be explored in this context.

\acknowledgments
The authors acknowledge the funding provided under projects AYA2006-07637 (Spanish Ministerio de 
Educaci\'on y Ciencia) and PCTIB2005GC3-03 (Conselleria d'Economia, 
Hisenda i Innovaci\'o of the Government of the Balearic Islands).



\begin{thebibliography}{37}
\expandafter\ifx\csname natexlab\endcsname\relax\def\natexlab#1{#1}\fi

\bibitem[{{Arregui} {et~al.}(2007){Arregui}, {Andries}, {Van Doorsselaere},
  {Goossens}, \& {Poedts}}]{Arregui07}
{Arregui}, I., {Andries}, J., {Van Doorsselaere}, T., {Goossens}, M., \&
  {Poedts}, S. 2007, \aap, 463, 333

\bibitem[{{Aschwanden} {et~al.}(1999){Aschwanden}, {Fletcher}, {Schrijver}, \&
  {Alexander}}]{Aschwanden99}
{Aschwanden}, M.~J., {Fletcher}, L., {Schrijver}, C.~J., \& {Alexander}, D.
  1999, \apj, 520, 880

\bibitem[{{Ballester}(2005)}]{Ballester05}
{Ballester}, J.~L. 2005, Space Science Reviews, 121, 105

\bibitem[{{Ballester}(2006)}]{Ballester06}
---. 2006, Royal Society of London Philosophical Transactions Series A, 364,
  405

\bibitem[{{Bommier} {et~al.}(1994){Bommier}, {Landi Degl'Innocenti}, {Leroy},
  \& {Sahal-Brechot}}]{Bommier94}
{Bommier}, V., {Landi Degl'Innocenti}, E., {Leroy}, J.-L., \& {Sahal-Brechot},
  S. 1994, \solphys, 154, 231

\bibitem[{{Bommier} \& {Leroy}(1998)}]{BL98}
{Bommier}, V. \& {Leroy}, J.~L. 1998, in Astronomical Society of the Pacific
  Conference Series, Vol. 150, IAU Colloq. 167: New Perspectives on Solar
  Prominences, ed. D.~F. {Webb}, B.~{Schmieder}, \& D.~M. {Rust}, 434

\bibitem[{{Carbonell} {et~al.}(2004){Carbonell}, {Oliver}, \&
  {Ballester}}]{Carbonell04}
{Carbonell}, M., {Oliver}, R., \& {Ballester}, J.~L. 2004, \aap, 415, 739

\bibitem[{{D{\'{\i}}az} {et~al.}(2002){D{\'{\i}}az}, {Oliver}, \&
  {Ballester}}]{diaz02}
{D{\'{\i}}az}, A.~J., {Oliver}, R., \& {Ballester}, J.~L. 2002, \apj, 580, 550

\bibitem[{{Edwin} \& {Roberts}(1983)}]{ER83}
{Edwin}, P.~M. \& {Roberts}, B. 1983, \solphys, 88, 179

\bibitem[{{Engvold}(1998)}]{Engvold98}
{Engvold}, O. 1998, in Astronomical Society of the Pacific Conference Series,
  Vol. 150, IAU Colloq. 167: New Perspectives on Solar Prominences, ed. D.~F.
  {Webb}, B.~{Schmieder}, \& D.~M. {Rust}, 23

\bibitem[{{Forteza} {et~al.}(2007){Forteza}, {Oliver}, {Ballester}, \&
  {Khodachenko}}]{Forteza07}
{Forteza}, P., {Oliver}, R., {Ballester}, J.~L., \& {Khodachenko}, M.~L. 2007,
  \aap, 461, 731

\bibitem[{{Goossens}(2008)}]{goossens08}
{Goossens}, M. 2008, in Waves \& Oscillations in the Solar Atmosphere: Heating
  and Magneto-Seismology, Proceedings of the International Astronomical Union,
  IAU Symposium, ed. R.~{Erd\'elyi} \& C.~A. {Mendoza-Brice\~no}, Vol. 247, 228

\bibitem[{{Goossens} {et~al.}(2006){Goossens}, {Andries}, \& {Arregui}}]{GAA06}
{Goossens}, M., {Andries}, J., \& {Arregui}, I. 2006, Royal Society of London
  Philosophical Transactions Series A, 364, 433

\bibitem[{{Goossens} {et~al.}(2008){Goossens}, {Arregui}, {Ballester}, \&
  {Wang}}]{GABW08}
{Goossens}, M., {Arregui}, I., {Ballester}, J.~L., \& {Wang}, T.~J. 2008, \aap,
  484, 851

\bibitem[{{Goossens} {et~al.}(1992){Goossens}, {Hollweg}, \&
  {Sakurai}}]{goossens92}
{Goossens}, M., {Hollweg}, J.~V., \& {Sakurai}, T. 1992, \solphys, 138, 233

\bibitem[{{Goossens} {et~al.}(1995){Goossens}, {Ruderman}, \&
  {Hollweg}}]{goossens95}
{Goossens}, M., {Ruderman}, M.~S., \& {Hollweg}, J.~V. 1995, \solphys, 157, 75

\bibitem[{{Hollweg} \& {Yang}(1988)}]{HY88}
{Hollweg}, J.~V. \& {Yang}, G. 1988, \jgr, 93, 5423

\bibitem[{{Leroy}(1980)}]{Leroy80}
{Leroy}, J.~L. 1980, in Japan-France Seminar on Solar Physics, ed.
  F.~{Moriyama} \& J.~C. {Henoux}, 155

\bibitem[{{Lin}(2004)}]{Linthesis}
{Lin}, Y. 2004, PhD thesis, University of Oslo, Norway

\bibitem[{{Lin} {et~al.}(2005){Lin}, {Engvold}, {Rouppe van der Voort}, {Wiik},
  \& {Berger}}]{Lin05}
{Lin}, Y., {Engvold}, O., {Rouppe van der Voort}, L., {Wiik}, J.~E., \&
  {Berger}, T.~E. 2005, \solphys, 226, 239

\bibitem[{{Lin} {et~al.}(2007){Lin}, {Engvold}, {Rouppe van der Voort}, \& {van
  Noort}}]{Lin07}
{Lin}, Y., {Engvold}, O., {Rouppe van der Voort}, L.~H.~M., \& {van Noort}, M.
  2007, \solphys, 246, 65

\bibitem[{{Martin} {et~al.}(2008){Martin}, {Lin}, \& {Engvold}}]{Martin08}
{Martin}, S.~F., {Lin}, Y., \& {Engvold}, O. 2008, \solphys, in press, DOI: 10.1007/s11207-008-9194-8

\bibitem[{{Molowny-Horas} {et~al.}(1999){Molowny-Horas}, {Wiehr}, {Balthasar},
  {Oliver}, \& {Ballester}}]{Molowny-Horas99}
{Molowny-Horas}, R., {Wiehr}, E., {Balthasar}, H., {Oliver}, R., \&
  {Ballester}, J.~L. 1999, in JOSO Annu. Rep., 1998, p. 126 - 127, 126--127

\bibitem[{{Nakariakov} {et~al.}(1999){Nakariakov}, {Ofman}, {DeLuca},
  {Roberts}, \& {Davila}}]{Nakariakov99}
{Nakariakov}, V.~M., {Ofman}, L., {DeLuca}, E.~E., {Roberts}, B., \& {Davila},
  J.~M. 1999, Science, 285, 862

\bibitem[{{Oliver} \& {Ballester}(2002)}]{OB02}
{Oliver}, R. \& {Ballester}, J.~L. 2002, \solphys, 206, 45

\bibitem[{{Ruderman} \& {Roberts}(2002)}]{RR02}
{Ruderman}, M.~S. \& {Roberts}, B. 2002, \apj, 577, 475

\bibitem[{{Sakurai} {et~al.}(1991){Sakurai}, {Goossens}, \&
  {Hollweg}}]{sakurai91}
{Sakurai}, T., {Goossens}, M., \& {Hollweg}, J.~V. 1991, \solphys, 133, 227

\bibitem[{{Sewell}(2005)}]{sewell05}
{Sewell}, G. 2005, The Numerical Solution of Ordinary and Partial Differential
  Equations (Wiley-Interscience)

\bibitem[{{Soler} {et~al.}(2007){Soler}, {Oliver}, \& {Ballester}}]{Soler07}
{Soler}, R., {Oliver}, R., \& {Ballester}, J.~L. 2007, \aap, 471, 1023

\bibitem[{{Soler} {et~al.}(2008){Soler}, {Oliver}, \& {Ballester}}]{Soler08}
---. 2008, \apj, in press

\bibitem[{{Terradas} {et~al.}(2005){Terradas}, {Carbonell}, {Oliver}, \&
  {Ballester}}]{Terradas05}
{Terradas}, J., {Carbonell}, M., {Oliver}, R., \& {Ballester}, J.~L. 2005,
  \aap, 434, 741

\bibitem[{{Terradas} {et~al.}(2002){Terradas}, {Molowny-Horas}, {Wiehr},
  {Balthasar}, {Oliver}, \& {Ballester}}]{Terradas02}
{Terradas}, J., {Molowny-Horas}, R., {Wiehr}, E., {Balthasar}, H., {Oliver},
  R., \& {Ballester}, J.~L. 2002, \aap, 393, 637

\bibitem[{{Terradas} {et~al.}(2001){Terradas}, {Oliver}, \&
  {Ballester}}]{Terradas01}
{Terradas}, J., {Oliver}, R., \& {Ballester}, J.~L. 2001, \aap, 378, 635

\bibitem[{{Terradas} {et~al.}(2006){Terradas}, {Oliver}, \&
  {Ballester}}]{TOB06}
---. 2006, \apj, 642, 533

\bibitem[{{Van Doorsselaere} {et~al.}(2004){Van Doorsselaere}, {Andries},
  {Poedts}, \& {Goossens}}]{tom04b}
{Van Doorsselaere}, T., {Andries}, J., {Poedts}, S., \& {Goossens}, M. 2004,
  \apj, 606, 1223

\bibitem[{{Yi} \& {Engvold}(1991)}]{YiEngvold91}
{Yi}, Z. \& {Engvold}, O. 1991, \solphys, 134, 275

\bibitem[{{Yi} {et~al.}(1991){Yi}, {Engvold}, \& {Keil}}]{Yi91}
{Yi}, Z., {Engvold}, O., \& {Keil}, S.~L. 1991, \solphys, 132, 63

\end{thebibliography}

\clearpage

\clearpage

\end{document}